\documentclass[aps,prl,floatfix,superscriptaddress,showpacs,twocolumn]{revtex4}
\usepackage{amsmath}
\usepackage{graphicx}

\begin{document}

\title{Effect of wavelength dependence of nonlinearity, gain, and dispersion 
in photonic crystal fiber amplifiers}
\author{A.\ Huttunen}
\affiliation{Department of Electrical and Communications Engineering, 
Laboratory of Computational Engineering, Helsinki University of Technology, 
FIN-02015 HUT, Finland}
\email{anu.huttunen@hut.fi}
\author{P.\ T\"orm\"a}
\affiliation{Department of Physics, Nanoscience Center, 
FIN-40014 University of Jyv\"askyl\"a, Finland}

\begin{abstract}
Photonic crystal fibers are used in fiber amplifiers and lasers because of 
the flexibility in the design of mode area and dispersion. However, these
quantities depend strongly on the wavelength. The wavelength dependence
of gain, nonlinearity and dispersion are investigated here by including 
the wavelength dependence explicitly in the nonlinear Schr\"odinger 
equation for photonic crystal fibers with varying periods and hole sizes. 
The effect of the wavelength dependence of each parameter is studied 
separately as well as combined. The wavelength dependence of the parameters 
is shown to create asymmetry to the spectrum and chirp, but to have a 
moderating effect on pulse broadening. 
The effect of including the wavelength dependence of nonlinearity in the 
simulations is demonstrated to be the most significant compared that 
of dispersion or gain.
\end{abstract}

\maketitle

\section{Introduction}

Photonic crystal fibers are a new class of optical fibers that have a 
periodic cladding~\cite{knight96,knight98}. 
Light can be confined to the core either by the band gap 
effect or by total internal reflection by average refractive index difference.
Photonic crystal fibers have many intriguing characteristics, 
for example, they can
be endlessly single mode, have extremely small or large mode areas and still
remain single mode. Also the dispersion properties of photonic crystal fibers
are very different from standard optical fiber.
The research of photonic crystal fiber lasers 
\cite{wadsworth00,furusawa01_el,sahu01,furusawa01,glas02,wadsworth03,limpert03,canning03,argyros04,mcneillie04,moenster04,mafi04,furusawa05,limpert05} 
and amplifiers 
\cite{price02,cucinotta03,hougaard03,limpert04,furusawa04,cucinotta04,li05,shirakawa05}
has been intense in the past few years. The aim has been two-fold.
First, the possibility of obtaining a small mode area has been exploited 
by designing high-gain amplifiers/lasers where the overlap
between the mode distribution and doped area is maximized.
On the other hand, the possibility of obtaining a large mode area
has been utilized for realizing high-power amplifiers/lasers with
low nonlinearity.

Small-mode area photonic crystal fibers generally have large
dispersion and nonlinearity that also depend strongly on the
wavelength~\cite{hainberger05}.
In this paper, we study the propagation of 200 fs pulses in 
high-gain small-mode area photonic crystal fiber amplifiers.
When short pulses are considered, the spectrum is wide and thus the 
wavelength dependence of the different parameters has a profound effect 
on the pulse propagation.

We compare the temporal and spectral widths, time-bandwidth products, 
and chirps of the pulses and amplification properties for different fiber 
geometries. 
We find out that the wavelength dependence of the parameters has
a substantial effect on the pulse properties after it has propagated 
a short distance in the fiber amplifier. The dispersion caused by the 
wavelength dependence of the nonlinearity counteracts the dispersion of 
the fiber and thus the pulses do not broaden as much as it is expected by 
considering constant parameter values for dispersion and nonlinearity.
On the other hand, the spectral broadening and chirping become
asymmetric when the wavelength dependence of the parameters is taken 
into account.
Also, the wavelength dependence of the dispersion parameters and gain is 
seen to influence the pulse quality less, compared to the important 
effect of wavelength dependence of nonlinearity, indicating that 
in some cases they could be approximated by constant values when 
simulating pulse propagation in photonic crystal fibers.
 
\section{Numerical methods}

The pulse propagation is studied with the optical nonlinear 
Schr\"odinger equation 
\begin{equation}
\begin{split}
\frac{\partial A}{\partial z}=&-\sum_{m=1}^4\frac{i^{m+1}}{m!}\beta_m\frac{\partial^mA}{\partial T^m}+\frac{g A}{2}\\
+&i\gamma\left(1+\frac{i}{\omega_0}\frac{\partial}{\partial T}\right ) A \int_{-\infty}^{\infty}R(T')\vert A(z,T-T')\vert^2 dT' ,
\label{schrodinger}
\end{split}
\end{equation}
which is simulated by the split-step Fourier method~\cite{agrawal}. 
The parameters for gain, nonlinearity and dispersion are taken to be either 
constants $g$, $\gamma$ and $\beta_m$ or wavelength dependent 
$g(\lambda)$, $\gamma(\lambda)$ and $\beta_m(\lambda)$, respectively.
The slowly varying envelope of the pulse is taken to be Gaussian
\begin{equation}
A(z=0,T)=\sqrt{P_0}e^{-T^2/(2T_0^2)},
\end{equation}
where $P_0$ is the peak power and $T_0$ is the pulse length.

The values of the dispersion and nonlinear parameters are calculated as
explained in Ref.~\cite{oma}.
The dispersion parameters $\beta_m$ are defined as
\begin{equation}
\beta_m=\left[\frac{d^m\beta}{d\omega^m}\right].
\label{beta}
\end{equation}
The mode-propagation constants $\beta$ as a function of the frequency $\omega$
were calculated with the full-vectorial plane wave method  
(the MIT Photonic bands software)~\cite{johnson01}.
The nonlinear parameter 
\begin{equation}
\gamma(\lambda)=\frac{2\pi}{\lambda}\frac{n_2}{A_{\rm eff}(\lambda)},
\label{gamma}
\end{equation}
is inversely proportional to the effective area,
which is calculated from the intensity distribution of the
eigenmode 
\begin{equation}
A_{\rm eff}(\lambda)=\frac{\left[ \int I(r)dr\right ] ^2 }{\int I^2(r)dr}
\end{equation}
and $n_2$ is the nonlinear-index coefficient $n_2=3\cdot 10^{-20}$m$^2$/W.

We consider an Erbium-doped fiber amplifier.
The gain as a function of wavelength is as in  Fig.~1 of
Ref.~\cite{cucinotta04} and it is approximated to be the same for 
all geometries since it has the least effect on the pulse propagation 
characteristics such as pulse broadening.
Also, according to Ref.~\cite{cucinotta03} the gain dependence on the
wavelength is strongly influenced by the emission and absorption cross 
section rather than by the photonic crystal fiber geometry.

\section{The studied fiber geometries and pulse properties}
\label{sec:geom}

\begin{figure}
\centering
\includegraphics[width=6cm]{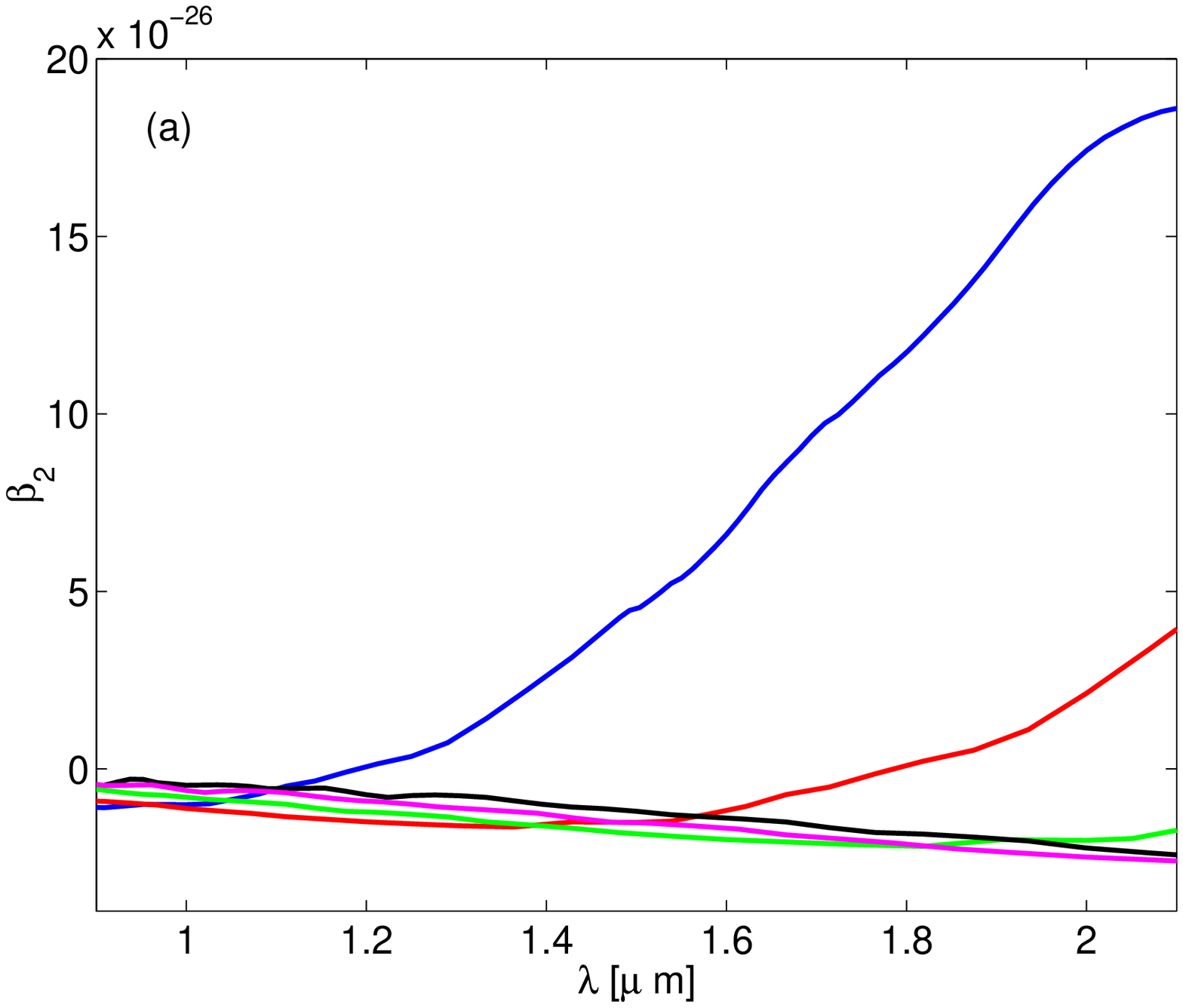}
\includegraphics[width=6cm]{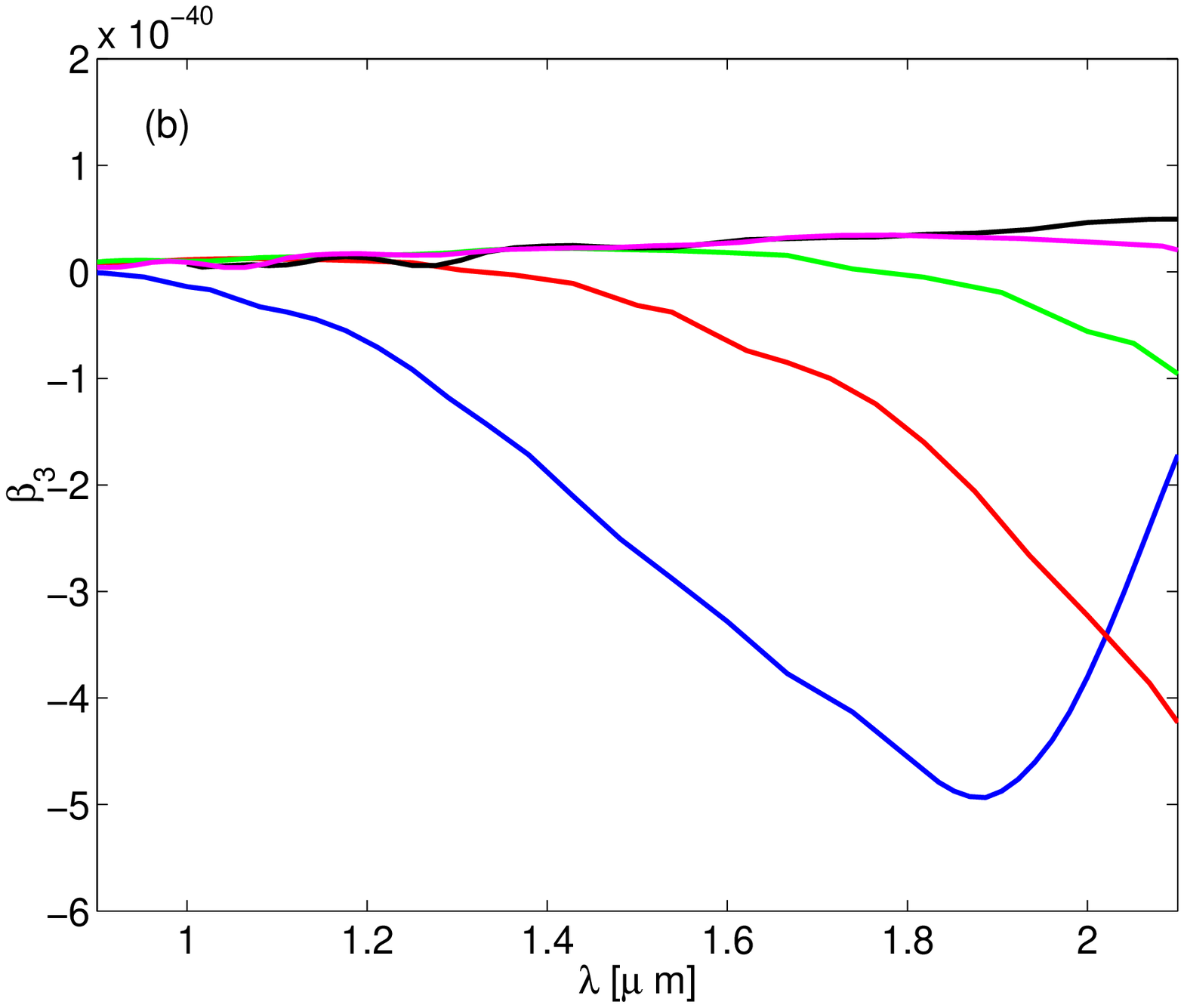}
\includegraphics[width=6cm]{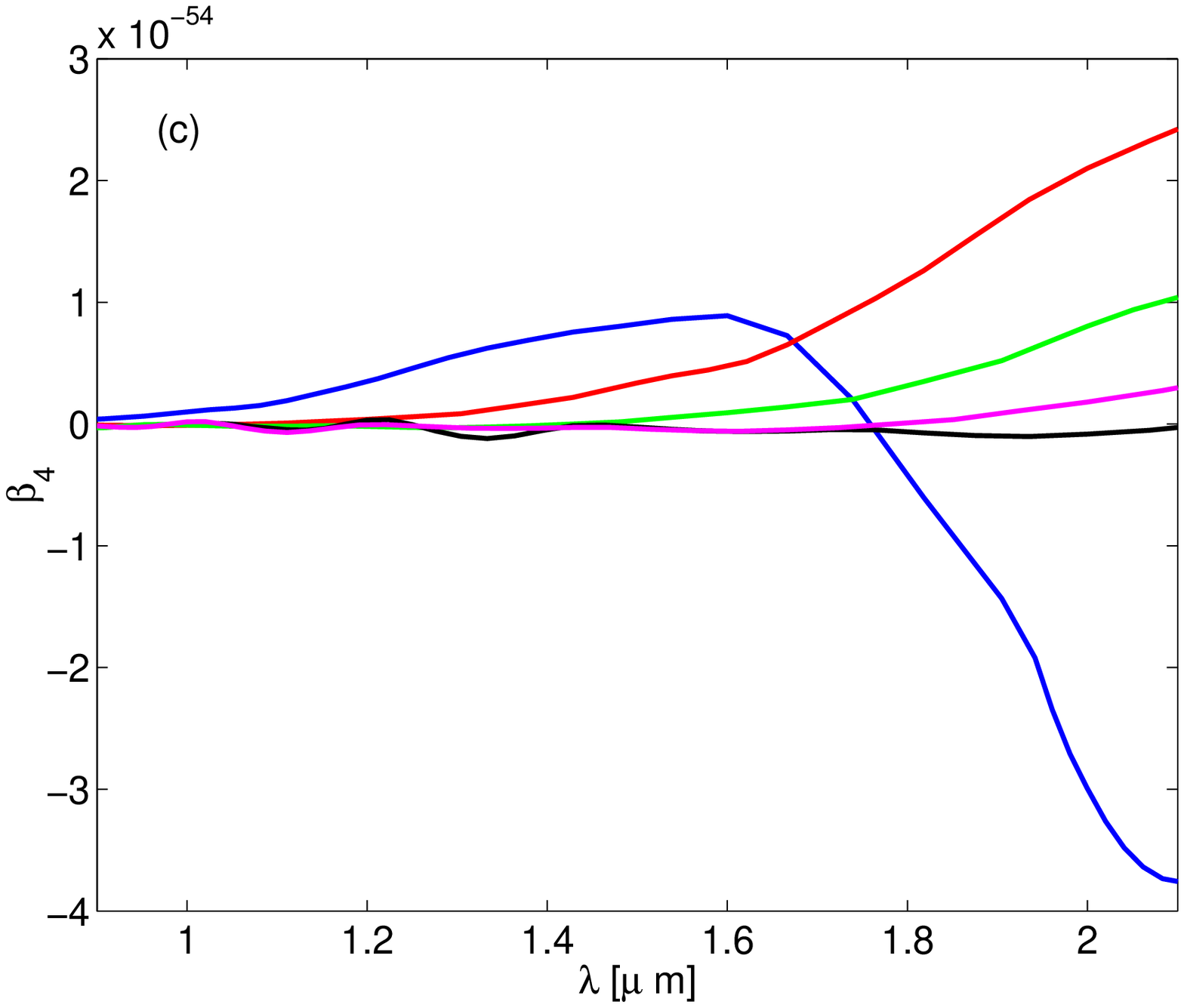}
\includegraphics[width=6cm]{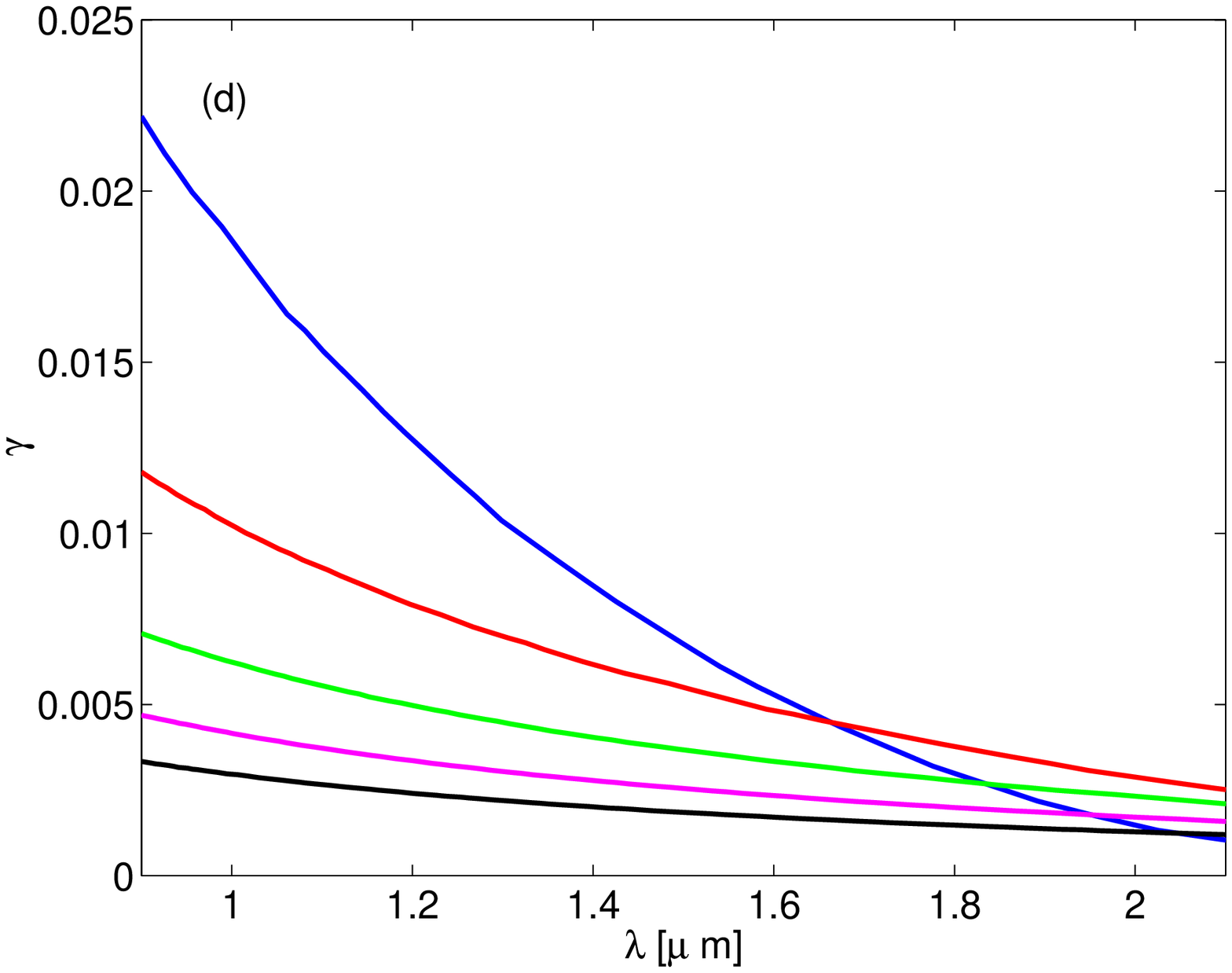}
\caption{Dispersion and nonlinear parameters
as a function of wavelength for a photonic crystal fiber with $d/P=0.3$
and periods $P=2$ $\mu$m (blue), $P=3$ $\mu$m (red), 
$P=4$ $\mu$m (green), $P=5$ $\mu$m (magenta),
and $P=6$ $\mu$m (black).}
\label{fig:coeffs}
\end{figure}

We investigate photonic crystal fibers with a triangular lattice of air holes 
in the cladding. The core is formed by a missing hole.  
We consider geometries with the hole diameter to period ratios
$d/P=$  0.2, 0.3, 0.4, 0.5, 0.6 and periods $P=2$, 3, 4, 5, 6 $\mu$m.
The geometries are those of interest in the research on high-gain 
efficiency photonic crystal fiber 
amplifiers~\cite{furusawa05,hougaard03,furusawa04,cucinotta04}.
The dispersion and nonlinear parameters [see Eqs.~(\ref{beta}) and 
(\ref{gamma})] are calculated as a function of the wavelength for all
the geometries. They are shown in Fig.~\ref{fig:coeffs} 
for the fiber geometry $d/P=0.3$.

The magnitude of the period affects the wavelength dependence of the $\beta_m$.
Regardless of $d/P$, for $P=2$ $\mu$m, the $\beta_m$ are strongly dependent 
on the wavelength while for $P=6$ $\mu$m the wavelength dependence is not as 
prominent.
The functional form of the $\beta_m$ as a function of the wavelength is 
very different for the different fiber geometries. For example, 
for $P=2$ $\mu$m, the value of $\beta_2$ in the center of
the considered wavelength range
is positive when $d/P$=0.2, 0.3, and 0.4, but negative when $d/P$=0.5 and 0.6,
which can affect the pulse propagation considerably.
The nonlinear coefficient $\gamma(\lambda)$, however, has a similar
form as a function of wavelength for all fiber geometries. 
The steepness of $\gamma(\lambda)$ increases when $P$ decreases or
$d/P$ increases.

The studied pulse has the length 200 fs, wavelength 1.55 $\mu$m, and peak 
power $P_0=$0.01 W. The propagation distance is 10 cm. The time and 
frequency axis are divided into $4096$ steps in the
split-step Fourier-method.

To demonstrate the effect of the wavelength dependence of $g(\lambda)$, 
$\gamma(\lambda)$, and $\beta_m(\lambda)$ in Eq.~(\ref{schrodinger}), 
the simulations for all the different geometries 
are performed twice: with wavelength dependent gain, nonlinearity and 
dispersion parameters  and, for comparison, 
with all these parameter values constant. The constant parameter values 
are determined at $\lambda=1.55$ $\mu$m.
The simulations for some of the geometries ($d/P=0.4$ and $d/P=0.5$) 
are repeated keeping the nonlinear parameter and gain wavelength
dependent, but approximating the the dispersion parameters with
constant values.
Also, to investigate the effect of the wavelength dependence of gain, the 
simulations are repeated with all other parameters constant
but including the wavelength dependence of gain.

\section{Comparison between simulations with constant and wavelength 
dependent gain, nonlinearity, and dispersion}

\begin{figure}
\centering
\includegraphics[width=6cm]{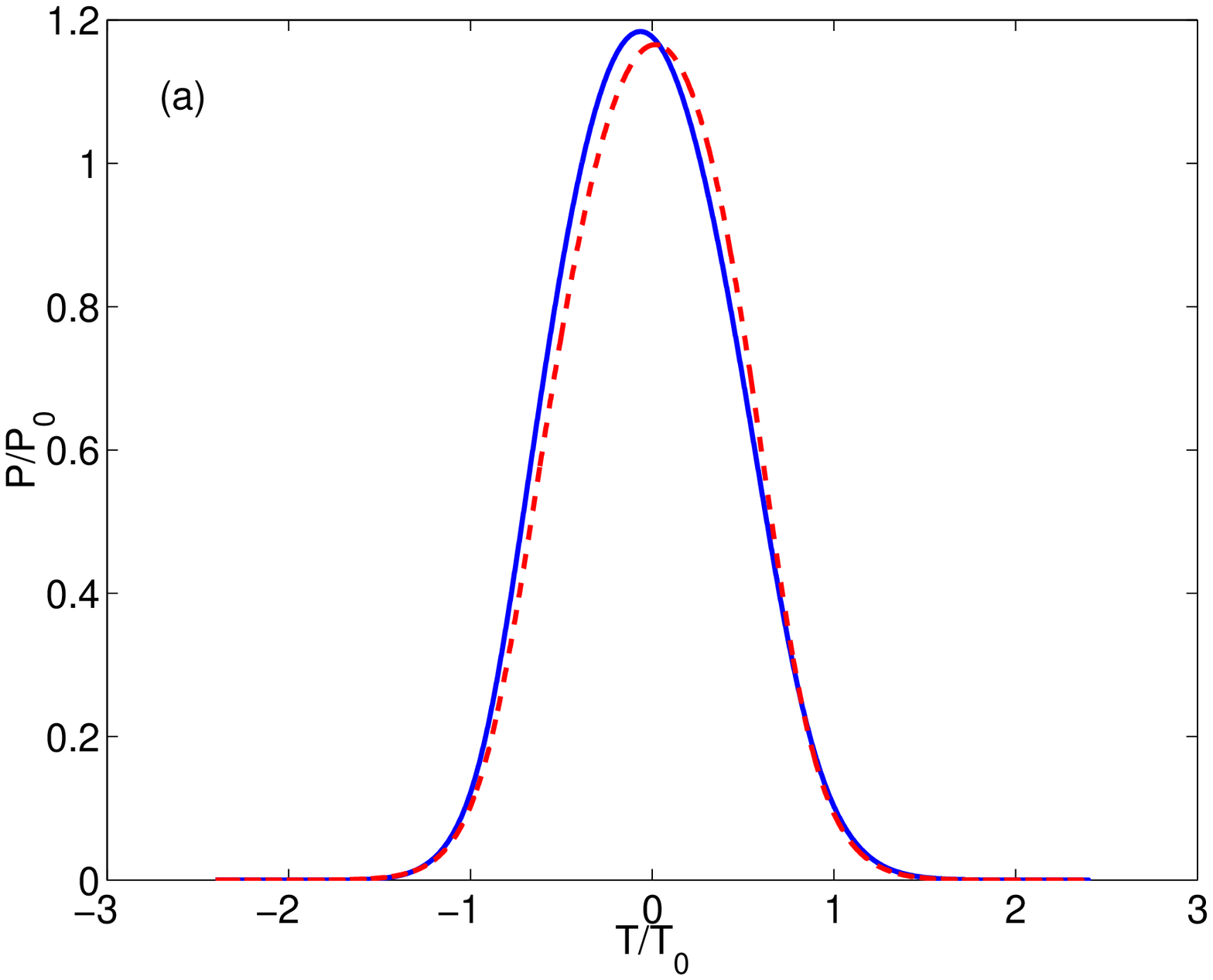}
\includegraphics[width=6cm]{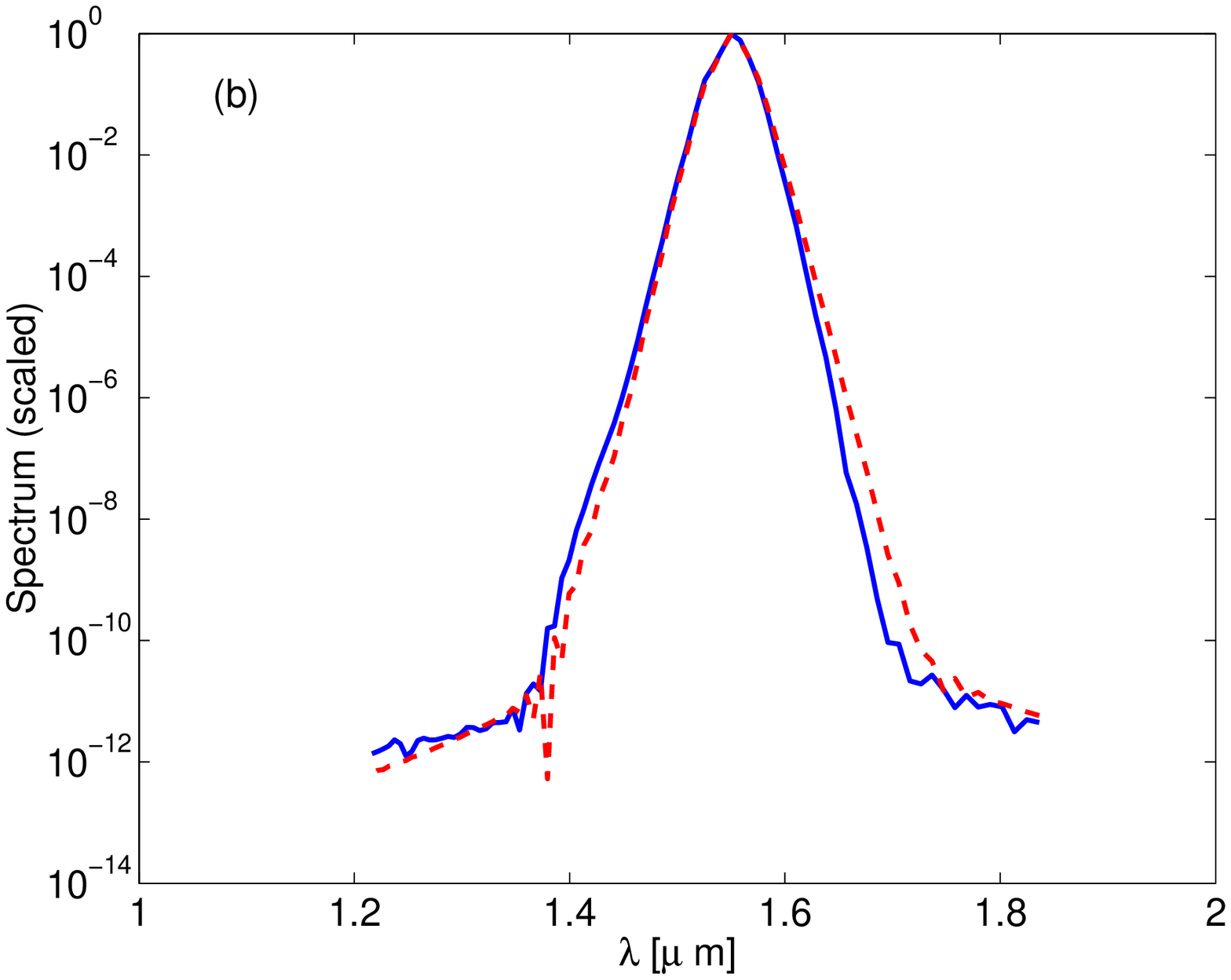}
\includegraphics[width=6cm]{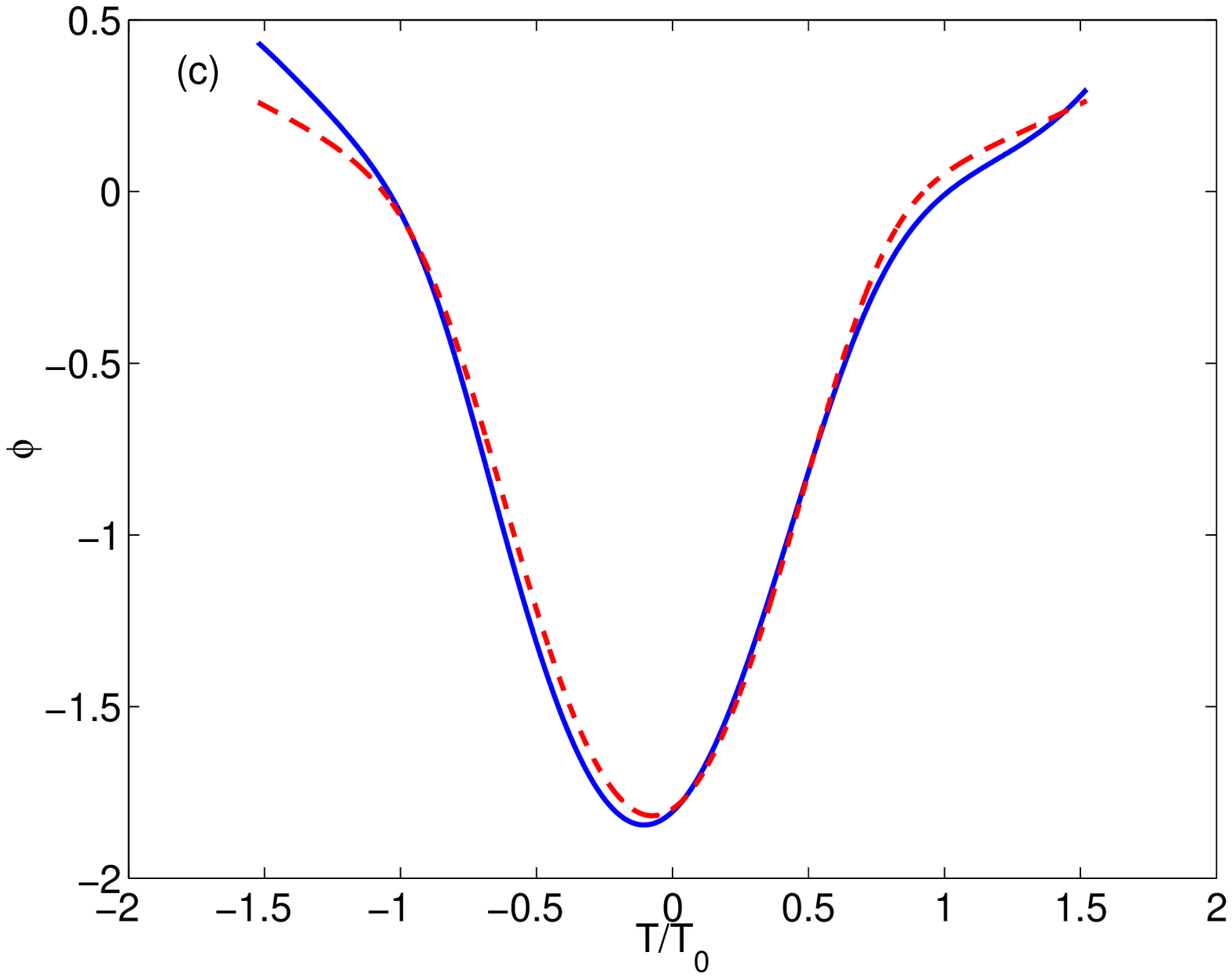}
\includegraphics[width=6cm]{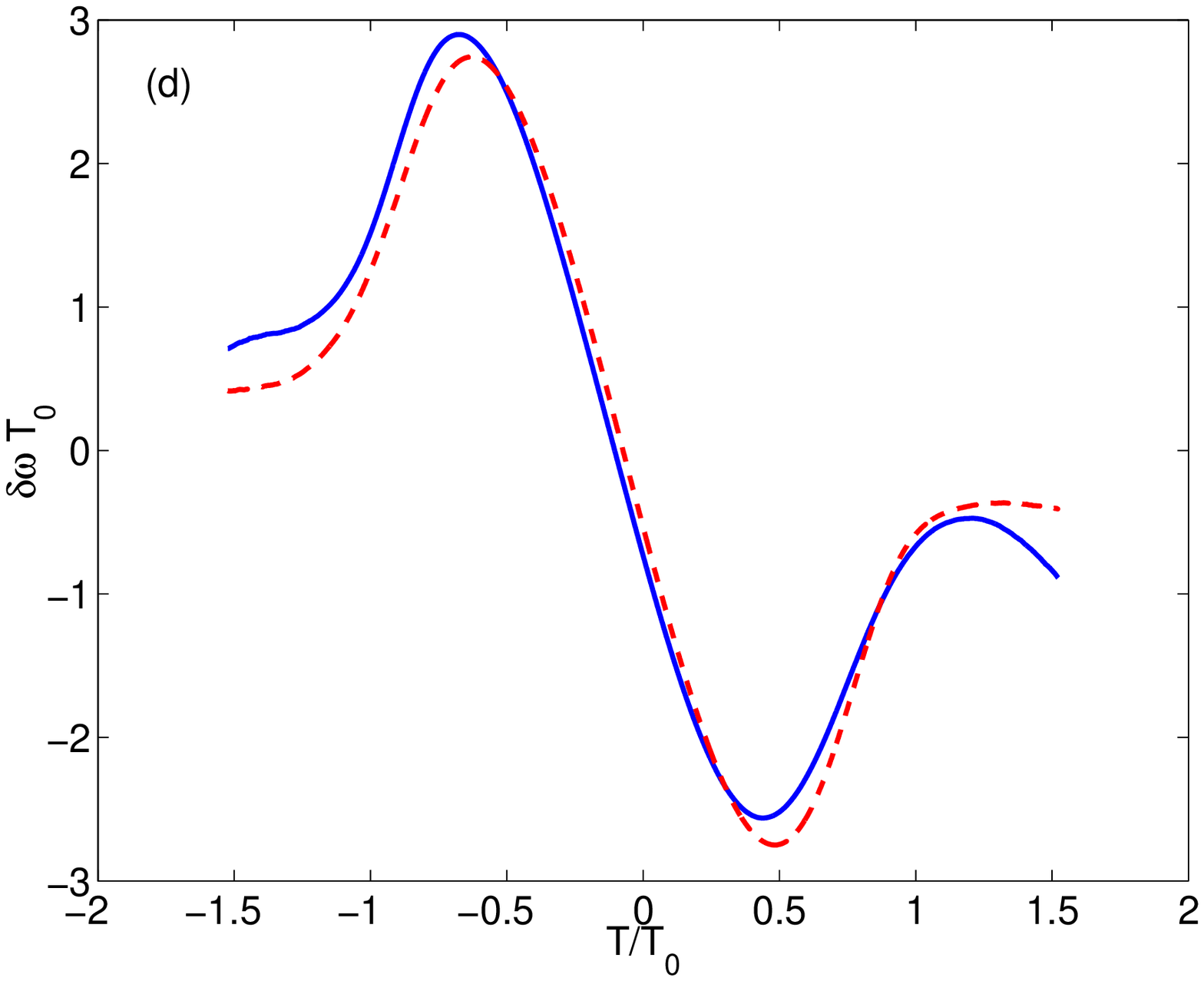}
\caption{(a) Pulse shape, (b) spectrum, (c) phase, and (d) chirp
after it has propagated 10 cm in the fiber amplifier.
The fiber geometry is $d/P=0.3$ and $P=3$ $\mu$m. 
Blue curves represent the simulation with wavelength dependent parameters
[$g(\lambda)$, $\gamma(\lambda)$, $\beta_m(\lambda)$]
and red dashed curves with constant parameters ($g$, $\gamma$, $\beta_m$).
The propagation distance is short compared to the total lengths of 
the fiber amplifiers (which can be several meters).}
\label{fig:shape}
\end{figure}

We compare the simulations made with wavelength dependent and constant 
gain, nonlinear and
dispersion parameters. After propagating 10 cm in the fiber amplifier, 
which is a short distance compared to actual amplifier lengths, 
there are clear differences for the two sets simulations.
The pulse shape, spectrum, phase, and chirp are shown in Fig.~\ref{fig:shape}
for one fiber geometry. The definitions of the phase of the pulse $\phi$ 
and frequency chirp $\delta\omega$ are
\begin{equation}
A(z,T)=\vert A(z,T)\vert e^{i\phi(z,T)}
\end{equation}
\begin{equation}
\delta\omega=-\frac{\partial\phi}{\partial T}.
\end{equation}
The chirp is a measure how much the instantaneous frequency changes across 
the pulse from the central frequency~\cite{agrawal}.
From Fig.~\ref{fig:shape} one can see that the self-steepening of the pulse
is larger for the simulation with constant $g$, $\gamma$, and $\beta_m$
than for wavelength dependent $g(\lambda)$, $\gamma(\lambda)$, and 
$\beta_m(\lambda)$. 
The spectrum of the pulse is more asymmetric when the wavelength dependence 
of the parameters is taken into account.
Also, the chirp of the simulation with constant parameters is symmetric 
whereas the chirp of the one with the wavelength dependent parameters 
is larger (in absolute magnitude) on the leading edge of the pulse than 
on the trailing edge.

\begin{figure}
\centering
\includegraphics[width=6cm]{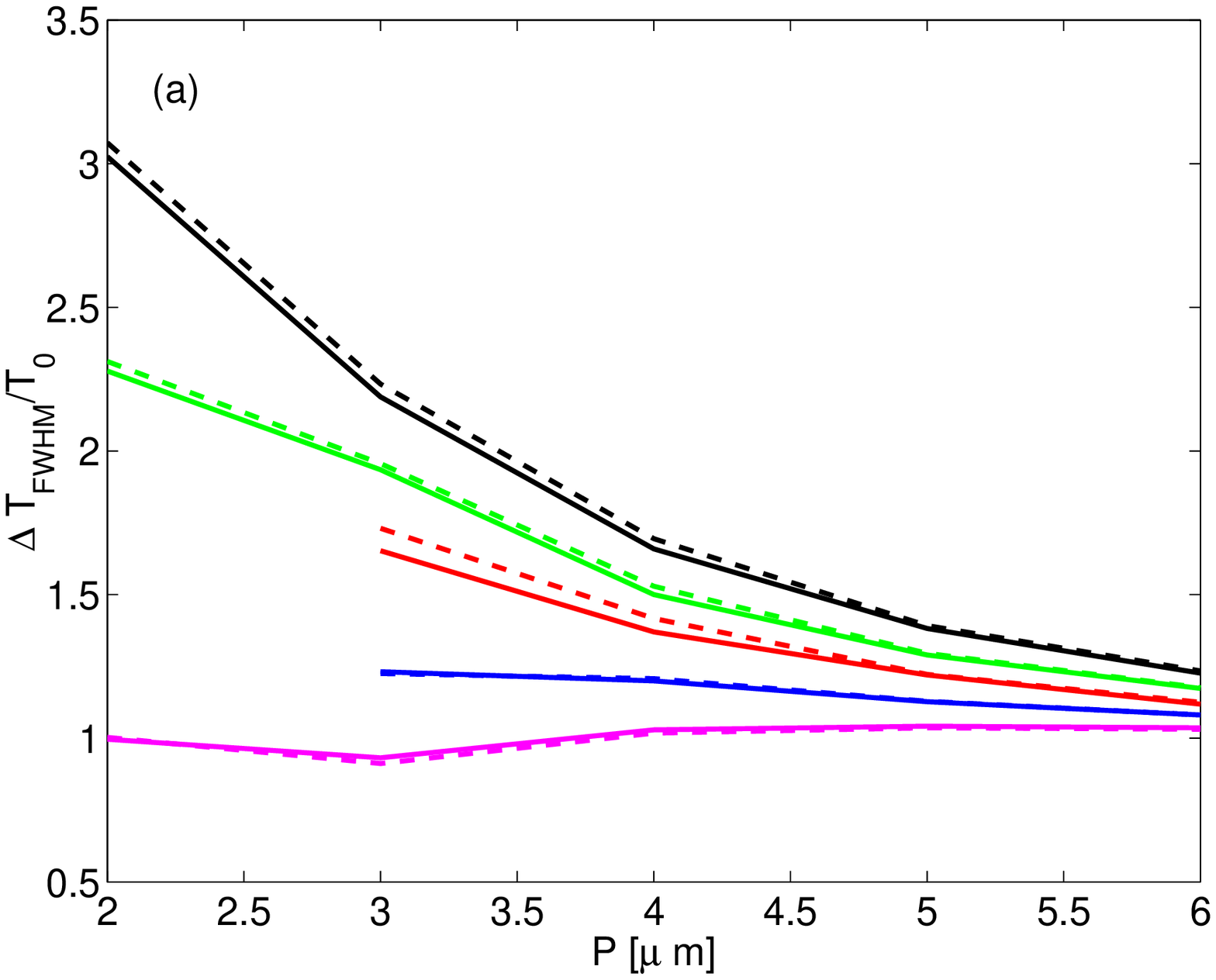}
\includegraphics[width=6cm]{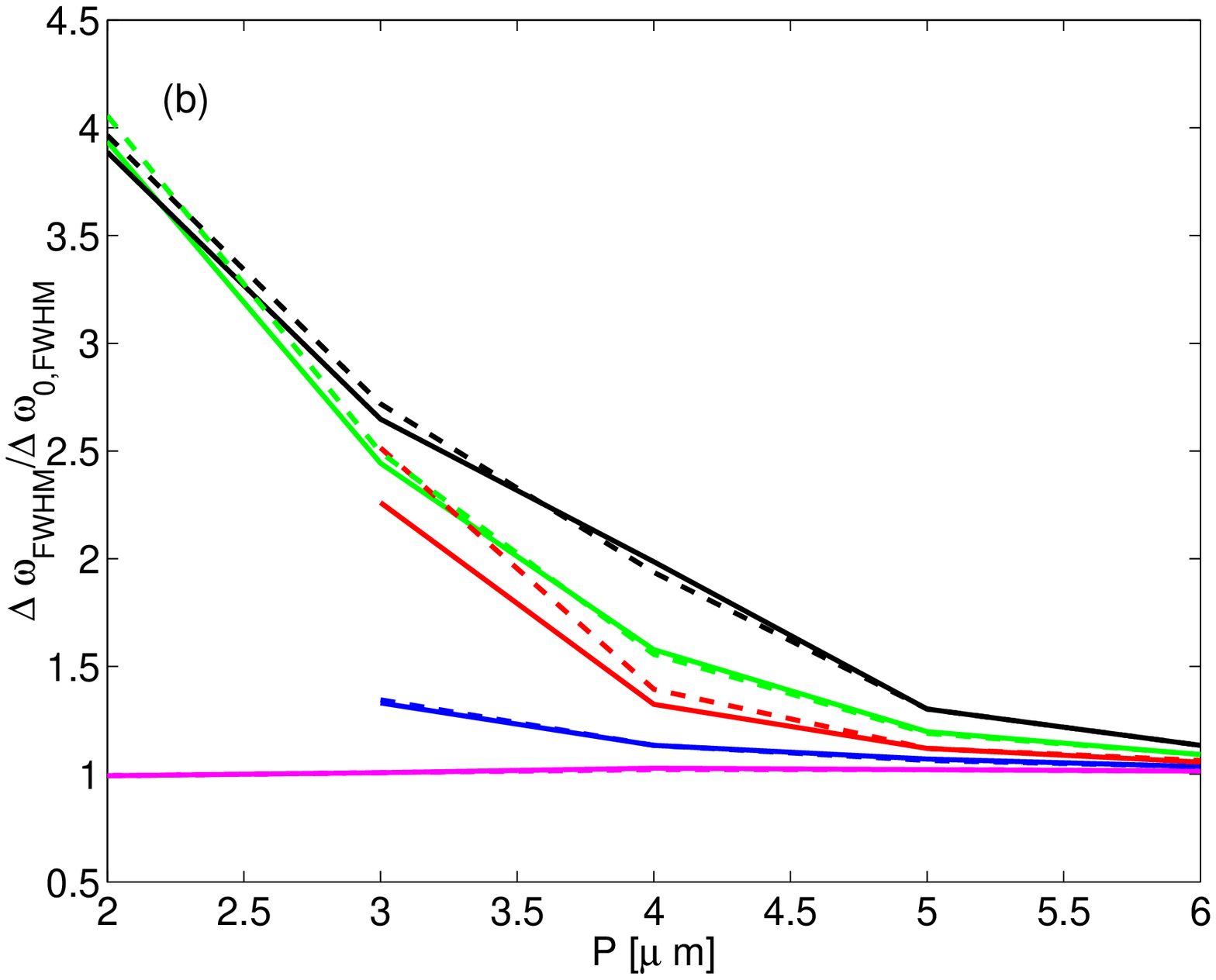}
\includegraphics[width=6cm]{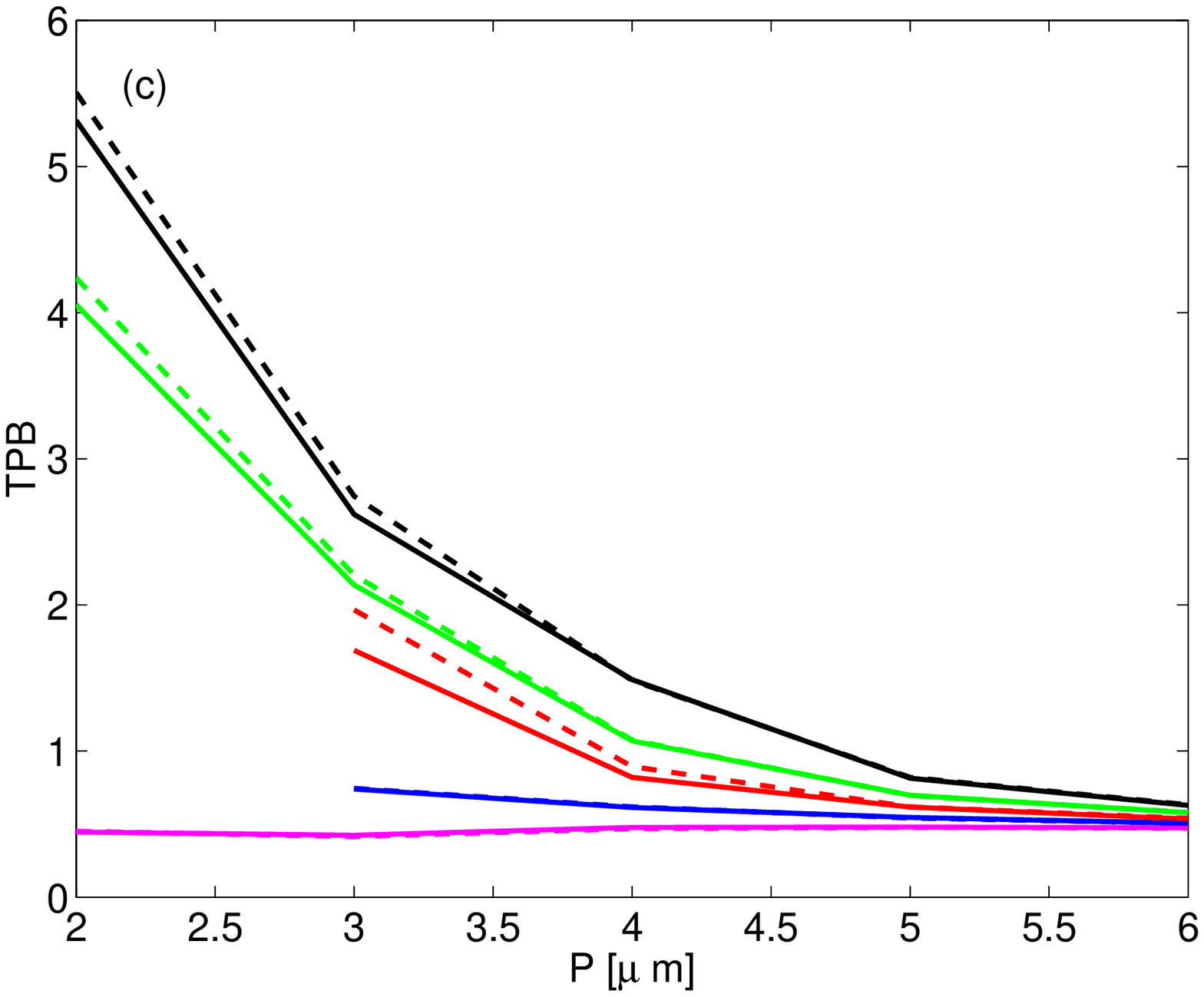}
\includegraphics[width=6cm]{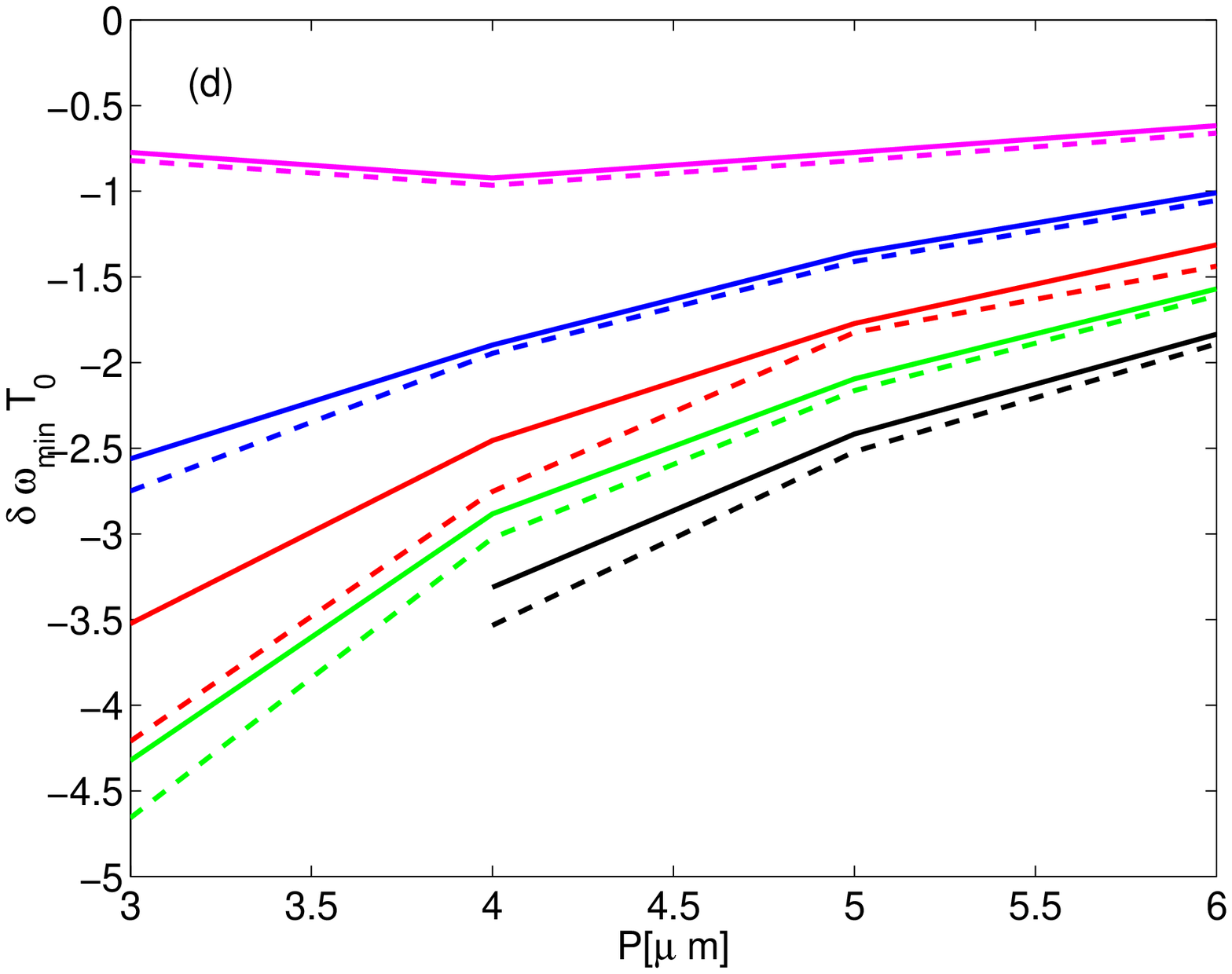}
\caption{(a) Temporal full-width at half-maximum, (b) spectral full-width 
at half-maximum, (c) time-bandwidth product, and (d) chirp
of the pulses after propagating 10 cm in the fiber amplifier
for the different fiber geometries: 
$d/P=0.2$ (magenta curves), $d/P=0.3$ (blue curves),
$d/P=0.4$ (red curves), $d/P=0.5$ (green curves), and $d/P=0.6$ (black curves)
as a function of the period $P$.
Solid curves represent the simulations for which the wavelength dependence of 
$g$, $\gamma$, and $\beta_m$ is taken into account. 
Dashed curves represent the simulations 
where these parameters are constant. 
The pulse widths are scaled by the width of the initial Gaussian pulse.
The chirp is the lowest value of the chirp 
[compare to Fig.~\ref{fig:shape}(d)].}
\label{fig:results}
\end{figure}

To characterize the pulses after the propagation in the amplifier,
we calculate the temporal and spectral widths, 
time-bandwidth products, and chirps of the pulses as well as their
amplification in the fiber. 
The temporal width is described by the full-width at half-maximum (FWHM)
and denoted by $\Delta T_{FWHM}$. We also calculated the root-mean square 
widths of the pulses, but they are not shown here since there was no 
additional information compared to the FWHM widths. 
The spectral width is calculated as the full-width at half-maximum 
$\Delta \omega_{FWHM}$. 
The pulse widths, the time-bandwidth products 
\begin{equation}
{\rm TBP}=\frac{\Delta T_{FWHM}\Delta \omega_{FWHM}}{2\pi}
\end{equation}
and the chirps are shown in Fig.~\ref{fig:results} as a function of 
the period for the different fiber geometries.
The chirp is indicated by the lowest value of the chirp 
[compare to Fig.~\ref{fig:shape}(d)].

The results for constant and wavelength dependent $g$, $\gamma$, and 
$\beta_m$ are different.
Pulses broaden less when the wavelength dependence of the parameters is 
taken into account, for all fiber geometries except $d/P=0.2$. 
This indicates that the wavelength dependence of 
the nonlinearity counteracts the temporal broadening induced by the dispersion.
As was discussed in Section \ref{sec:geom}, the form of the nonlinearity as 
a function of the wavelength is similar for all the fiber geometries,
only the steepness of $\gamma(\lambda)$ decreases for decreasing $d/P$,
whereas the form of the dispersion as a function of the wavelength 
is completely different for all fiber geometries.
From Fig.~\ref{fig:results}(a) one can see that the temporal 
broadening of the pulses is reduced for all fiber geometries,
only for $d/P=0.2$ the effect has vanished, due to the slight steepness
of the nonlinearity curve.

The magnitude of the chirping is not strongly affected when the
wavelength dependence of the parameters is included, but it shows 
a clear asymmetry that can be seen in Figs.~\ref{fig:shape}(c) 
and \ref{fig:results}(d). The chirp magnitudes for the geometries 
with $P=2$ $\mu$m, $d/P=0.2,0.3,0.4,0.5,0.6$ and for $P=3$ $\mu$m,
$d/P=0.6$, are not shown in
Fig.~\ref{fig:results}(d) since the chirps are linear across the total 
length of the pulse and thus exhibit no relevant minima.

It can be seen from Fig.~\ref{fig:results} that 
the pulses broaden more when $d/P$ increases or $P$ decreases, because
the nonlinear and dispersion parameters increase 
with increasing $d/P$ and/or decreasing $P$.
For fiber geometries that have large periods, the $d/P$ has less effect 
on the pulse broadening, because the dispersion and nonlinearity of large
$d/P$ geometries do not depend strongly on $P$.
The geometries with small $P$ and $d/P$ are not suitable for short pulse
amplification since the broadening of the pulses is so strong.
Also, the chirping lowers the quality of the pulses and effects 
dispersion compensation schemes.

\begin{figure}
\centering
\includegraphics[width=6cm]{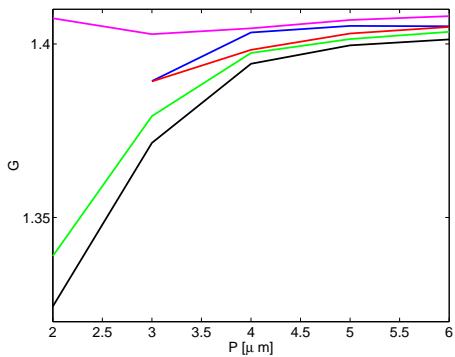}
\caption{The amplification of the pulses after propagating 10 cm 
in the fiber amplifier for the different fiber geometries: 
$d/P=0.2$ (magenta curve), $d/P=0.3$ (blue curve),
$d/P=0.4$ (red curve), $d/P=0.5$ (green curve), and $d/P=0.6$ (black curve)
as a function of the period $P$. }
\label{fig:amp}
\end{figure}

The amplification in the 10 cm of the fiber amplifier was calculated as
\begin{equation}
G=\frac{\int_{-\infty}^{\infty}\vert A(z,t)\vert ^2dT}{\int_{-\infty}^{\infty}\vert A(z=0,t)\vert ^2dT}
\end{equation}
and it is shown in Fig.~\ref{fig:amp}. Here the wavelength dependence 
of the parameters is taken into account.
The fibers with small $P$ and/or large $d/P$ are amplified less since 
the spectrum of the pulses broadens heavily and not all frequency components 
get amplified since the Erbium gain spectrum is so limited.
This indicates that the effects of pulse spectrum broadening have to be 
taken into account when high-gain fiber amplifiers are designed.

\begin{figure}
\centering
\includegraphics[width=6cm]{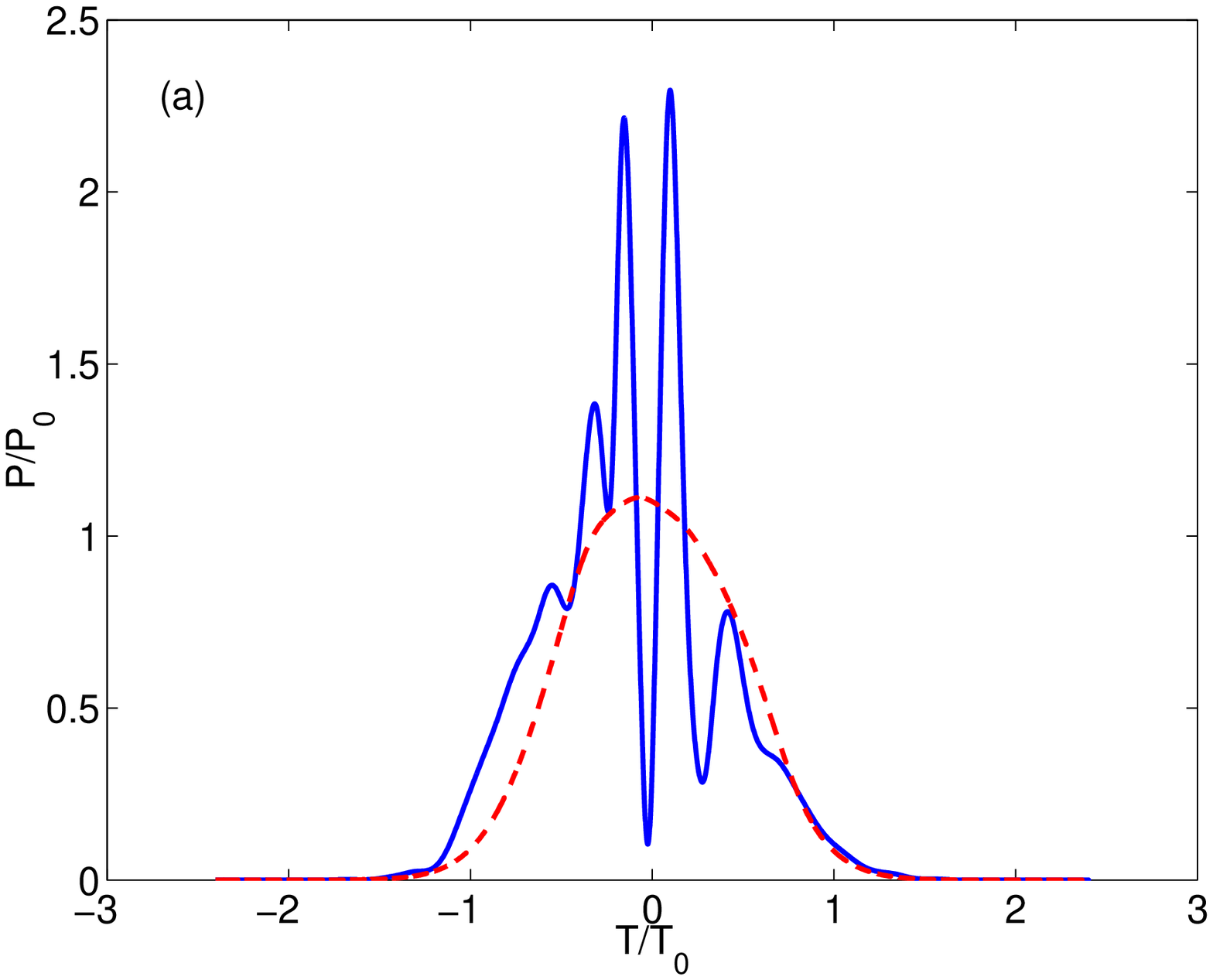}
\includegraphics[width=6cm]{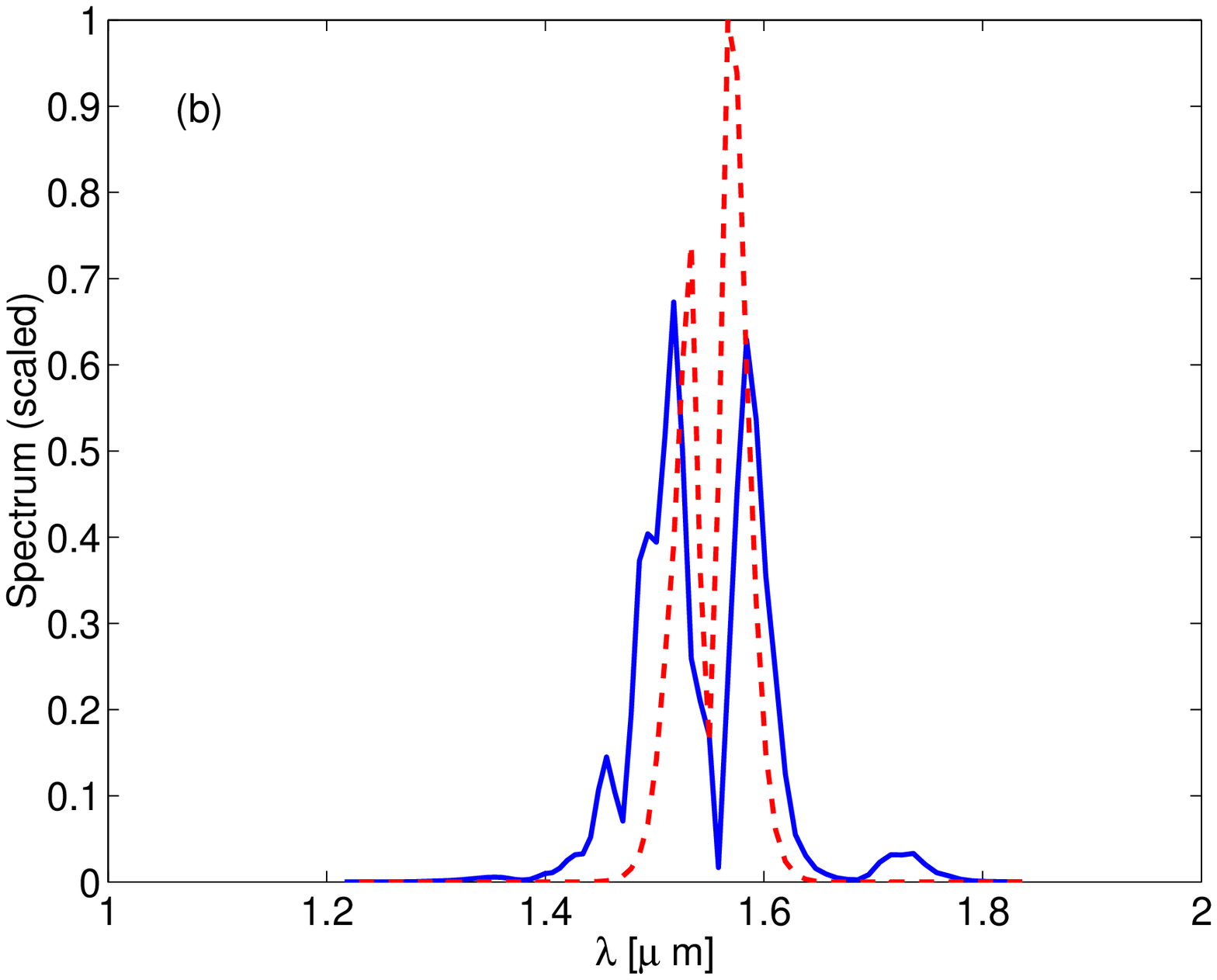}
\caption{(a) Pulse shape and (b) spectrum after propagating 9 cm in
the fiber amplifier with geometry $d/P=0.4$ 
and $P=2$ $\mu$m. Blue and red dashed curves represent the simulation 
with wavelength dependent and constant parameters $g$, $\gamma$, $\beta_m$, 
respectively.}
\label{fig:wave_breaking}
\end{figure}

The results of two fiber geometries 
($d/P=0.3$, $P=2$ $\mu$m and $d/P=0.4$, $P=2$ $\mu$m) 
are not shown in Figs.~\ref{fig:results} and \ref{fig:amp}.
These fiber geometries have very large third order dispersion which
results in a complicated pulse shape (see Fig.~\ref{fig:wave_breaking}).
The behavior of these pulses
is completely different for the two simulations with constant and 
wavelength dependent parameters 
(compare the blue and red dashed curves in Fig.~\ref{fig:wave_breaking}). 
Generally, when $P$ is small small, the parameters $\beta$ and $\gamma$
depend strongly on the wavelength, 
and they cannot be approximated by constant values.

\section{Comparison of the impact of the wavelength dependence of
the different fiber parameters}

\begin{figure}
\centering
\includegraphics[width=6cm]{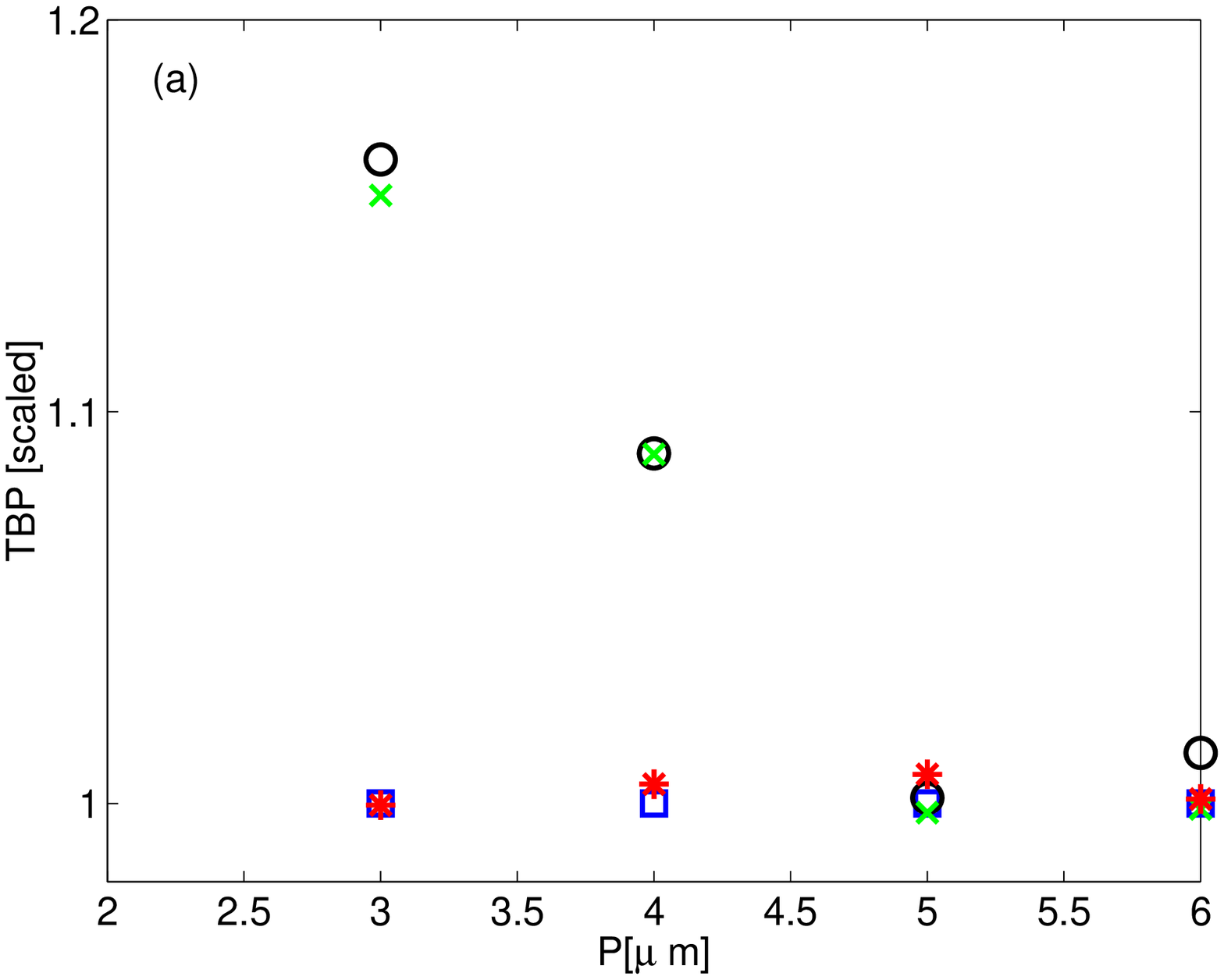}
\includegraphics[width=6cm]{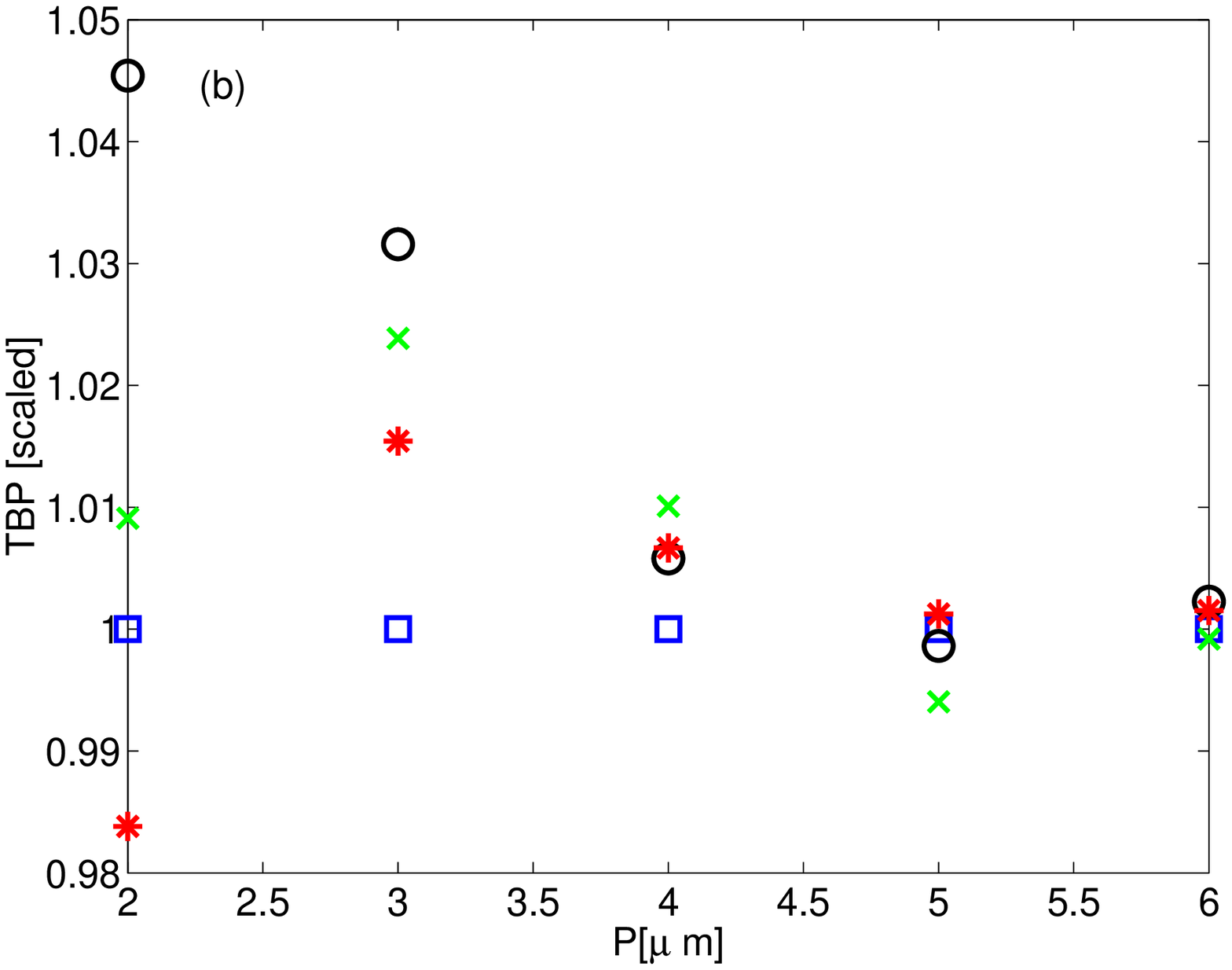}
\caption{Time-bandwidth products for geometries $d/P=0.4$ (a) and $d/P=0.5$ (b)
as a function of the period for four different sets of simulations.
Blue squares denote simulations with all parameters wavelength dependent and
red stars denote simulations where dispersion constants were approximated 
with constant values. Black circles denote simulations with all 
parameters constant and green crosses denote simulations where 
only gain is wavelength dependent.
The TBP values are scaled to the results from the simulations with 
all parameters wavelength dependent for the corresponding fiber geometry.}
\label{fig:comp}
\end{figure}

In order to compare the effect of including the wavelength dependence 
of the different fiber parameters on the pulse characteristics, 
we calculated the pulse propagation for the fiber geometries 
$d/P=0.4$ and $d/P=0.5$ with two new sets of simulations.
Firstly, we approximated the dispersion parameters $\beta_m$ with constant 
values, and secondly, we approximated all other parameters, except gain,
with constant values.
The time-bandwidth products are shown in Fig.~\ref{fig:comp}. 
From Fig.~\ref{fig:comp}(a) one can see that the results for the simulations 
where all the parameters were wavelength dependent (blue squares) and where 
dispersion parameters $\beta_m$ were approximated by constant values 
(red stars) are very close, compared to the simulations with all parameters 
constant (black circles). This indicates that the 
wavelength dependence of the dispersion coefficients $\beta_m$ does
not have as significant effect to the pulse as the 
wavelength dependence of the nonlinearity coefficient.
For the fiber geometries in Fig.~\ref{fig:comp}(b) the wavelength 
dependence of the dispersion has a larger impact on the pulse characteristics, 
since the pulse spectrum for the geometries with larger $d/P$ is wider.

To determine the effect of the wavelength dependence of gain,
we calculated the pulse propagation when only the wavelength dependence 
of gain was taken into account but all other parameters ($\beta_m$ and 
$\gamma$) were constant.
The time-bandwidth products are shown in Fig.~\ref{fig:comp}.
One can see in Fig.~\ref{fig:comp}(a) that the simulation where 
the wavelength dependence of gain is taken into account (green crosses) 
does not change the pulse properties 
in a significant amount compared to the simulation where all the parameters 
were constant (black circles). Again in Fig.~\ref{fig:comp}(b) the effect
of the wavelength dependence of gain is more important due to the 
extensive pulse broadening. 
This implies that the wavelength dependence of gain does not have a 
significant effect to the pulse propagation except in the case where 
the spectral broadening of the pulse is excessive.
For the fiber geometry with $d/P=0.5$ and $P=2$ $\mu$m, 
it is very important to include the wavelength dependence of gain in 
the simulation since the pulse broadens out of the gain spectrum of Erbium.

\section{Conclusions}

We have studied pulse propagation in high-gain efficiency photonic crystal 
fiber amplifiers with varying periods and hole sizes. 
We took into account the wavelength dependence of the 
fiber parameters for dispersion, nonlinearity, and gain. 
The wavelength dependence of the fiber parameters has a significant effect 
on the temporal and spectral width of the pulse. The pulses were shown 
to broaden less for most fiber geometries when the wavelength dependence 
of the parameters was taken into account, 
indicating that the wavelength dependence of the
nonlinearity counteracts the dispersion of the fibers.
The spectral width and chirp
showed asymmetry after propagating a short distance in the fiber amplifier.
This could affect for example pulse compression or dispersion compensation 
schemes. Although the changes in the pulse properties shown here were
qualitatively small, they are important since the propagation 
distance was short compared to actual amplifier lengths.
The wavelength dependence of dispersion and gain was shown not
to have as profound effect on the pulse quality as 
the wavelength dependence of nonlinearity. 
However, the wavelength dependence of all fiber parameters
have to be included in the simulations when the pulse
spectrum broadens heavily.

\section*{Acknowledgments}
We thank Emil Aaltonen foundation and the Academy of Finland
for support (Project Nos. 53903, 205454).

\end{document}